\documentclass{article}

\usepackage{microtype}
\usepackage{graphicx}
\usepackage{subfigure}
\usepackage{booktabs} % for professional tables
\usepackage{amsmath} % for equations
\usepackage{dirtytalk}
\graphicspath{{img/}}

\usepackage{xcolor}
\definecolor{topic_color}{RGB}{0, 150, 0}
\newcommand\hidden[1]{}
\newcommand\error[1]{}
\newcommand\topic[1]{}

\usepackage{hyperref} 

\usepackage[accepted]{icml2019}

\begin{document}

\twocolumn[
\title{Command Line Interface Risk Modeling}
\date{\vspace{-0.2in}}
\maketitle

%//////////////////// Author information ////////////////////////

% Include the name, group, and email of the author.
% Use the standard \icmlauthor{Name, Group, email@email.com}{}\\
% Do not remove the curly brackets {} at the end.

% Equal contribution.
% Adding "equal" to the second bracket will create an * meaning equal contribution.
% If equal contribution is used, read further instructions after square bracket ].
% Affiliations will be numbered in order of appearance
\icmlsetsymbol{equal}{*}

\begin{icmlauthorlist}
\icmlauthor{Dr~Anthony~L~Faulds, Microsoft, tonyfaulds@microsoft.com, https://orcid.org/0000-0002-7209-2472}{}
\end{icmlauthorlist}
\vspace{0.4in}

% Grammarly Score 83 6/3/2022
% Grammarly Score 86 6/21/2022
% Grammarly Score 98 6/23/2022
% Grammarly Score 99 7/11/2022
% Grammarly Score 99 8/29/2022

\begin{abstract}
Protecting sensitive data is an essential part of security in cloud computing. However, only specific privileged individuals have access to view or interact with this data; therefore, it is unscalable to depend on these individuals also to maintain the software. A solution to this is to allow non-privileged individuals access to maintain these systems but mask sensitive information from egressing. To this end, we have created a machine-learning model to predict and redact fields with sensitive data. This work concentrates on Azure PowerShell, showing how it applies to other command-line interfaces and APIs. Using the F5-score as a weighted metric, we demonstrate different transformation techniques to map this problem from an unknown field to the well-researched area of natural language processing.
\end{abstract}
\vspace{0.4in}
]

% This command creates the footnote for the equal contribution*
% If you need to mention equal contribution uncomment the next line
%	\printAffiliationsAndNotice{\icmlEqualContribution} % Uncomment for equal contribution
\clearpage

\section{Introduction}
\topic{introduction of what}
\topic{Description of Sovereign}
Sovereign Cloud is a unique offering by Microsoft that allows customers access to the features of Azure Cloud but with isolated physical resource infrastructure. Some Sovereign Clouds have varying levels of internet access depending on the customer's risk aversion. Each customer defines the requirements of a user to gain access to the system. These requirements can range from receiving certification to having government clearance. Users who fulfill the appropriate requirements and have administrator-level access are called operators. Maintaining the Azure Cloud requires directly responsible individuals (DRI) who can support and debug issues related to different Azure services.

\topic{Define sovereign people}
Unfortunately, the number of individuals that are both a DRI, have Azure expertise, and an operator, have the necessary credentials is a small group of people. This group of people is difficult to hire, train and grow because clearances can take months, and learning Azure enough to support a systems infrastructure can take months. One way to solve this is to separate the operator and DRI roles. As a result, DRIs can increase as Microsoft hires more developers in the Azure infrastructure. In addition, DRIs can interact directly with the Sovereign Cloud, but sensitive information must be redacted from their interactions. To solve the issue of redacting sensitive information, the Command And Query (CNQ) team developed a command-line interface (CLI) to allow DRIs to send commands to the Sovereign Cloud while using a combination of ML and operator feedback to redact sensitive information returned in CLI responses. This solution enables the number of DRIs to scale without requiring the number of operators to increase at the same rate or requiring that DRIs also be operators.

\topic{Shell commands}
PowerShell is a CLI that enables scripting and automation. Azure Cloud has created PowerShell commands that allocate resources, query resources, connect to resources, deletes resources, and more.

\topic{Sensitive}
Sensitive information is data in the Sovereign Cloud or how to connect to the data, connection strings, and passwords. Each Sovereign Cloud customer has its definition of sensitive information, but some pieces of data are universally sensitive across all customers.

\topic{modeling}
We show how we can map this problem to more widely researched NLP problems. The result is that each feature becomes a document with words inside that document. CLI risk prediction is a unique problem, and this work explores different ways to map the CLI command and its response to machine learning problems. Although we show how to map CLI data to documents, similar to the data sets in the NLP modeling space, CLI is different because it does not hold to traditional grammar rules and requires some brevity. For example, variable names and commands are long enough to be self-describing but shorter than it is context compared to a typical sentence.

\topic{What this work introduces}
Although this technique is specific to Powershell, the same process can apply to other CLIs and APIs. For example, changing from binary class to multiclass classifier can map this problem to determine various types of Personally Identifiable Information (PII) within command responses. In addition, we can adjust from predicting the probability of sensitive data to predicting the likelihood that a command can cause a system outage.

\section{Problem Definition}
\topic{define problem}
In security, it is crucial to constrain the access or use of sensitive information. This information should have limited transmission or visibility except for those who are allowed to see this data. Masking this data from the view of non-qualified individuals is one approach to security. It will enable non-qualified individuals or systems to maintain secure systems but limits the information seen.

\topic{sensitive information}
For Sovereign Cloud, most clients have similar definitions of sensitive information. For example, Key Vault information is considered sensitive information. Everything from passwords, connection strings, and certificates are examples of the many things a Key Vault may store. PowerShell has commands that can create resources that respond with connection string information and also have commands to fetch these fields specifically.

\topic{future}
As Sovereign Cloud customers evolve, things like the server names, geographical locations, or PII included in the system will fall under sensitive information in the future.

\topic{mapping from a domain to domain}
For this work, we want to map the PowerShell command-line redaction to a well-known problem. Since there are no numbers, it seems best to map this work to an NLP problem. Furthermore, ML models predict sensitivity per field to allow fine-grain resolution for redacting fields. Therefore a single PowerShell command is mapped to multiple predictions, one per response.

\section{Related Work}
\topic{Intro}
There are many approaches to NLP problems. Even early approaches converted words into numbers, or vectors for numerical analysis \cite{luhn1957statistical} \cite{teller2000speech} \cite{jones2004statistical}. For this work, we compare several of those to show the best combination.

\topic{bag of words}
The bag of words (BOW) method is a simplified text representation \cite{joachims1998text} \cite{zhang2010understanding}. It ignores word order and grammar and stores the count of words as a feature. The word counts can use all vocabulary in the training data set or a specified dictionary. Next, BOW converts a string of words into an integer vector with counts of the word in the document.

\topic{TF-IDF}
Term frequency-inverse document frequency (TF-IDF) extends BOW from word count to include information about how many words are in a document and how frequently the words are found across documents \cite{robertson2004understanding} \cite{vinokourov2002inferring} \cite{rajaraman2011mining}
\cite{lavelli2004distributional}. Beel, Gipp, Langer, and Breitinger have documented the popularity and usefulness of this algorithm \cite{beel2016paper}. TF-IDF is a combination of two components, TF and IDF. For TF, this is the ratio of the count of a single word in a document over the number of words in the document. Closer to 1.0 indicates how prevalent the word is in a document compared to other terms in that document. IDF is the inverse of how dominant a word is in all documents. For example, a word like \say{the} being in all documents would give it a unary IDF of 1, whereas \say{contemplate} being used in 1 of 1000 documents would give it a unary IDF of 1000. Multiplying these two numbers together results in TD-IDF, a combination of word frequency relative to document length and inverse popularity of the word. This combination results in large TD-IDF values only when a word is rare or used a lot in a single document. TF-IDF is used as inputs or features to different techniques, from clustering to visual search of videos \cite{moon2013survey} \cite{sivic2008efficient} \cite{ramos2003using}.

\topic{Word Embedding}
Another approach that helps aid in understanding similar words is word embedding. Mikolov, Chen, Corrado, and Dean developed a method of using neural networks to project words into a vector space \cite{mikolov2013distributed} \cite{mikolov2013efficient}. This technique transforms words into a real-value vector in which words close in vector distance have similar meanings. For example, BOW and TF-IDF treat \say{secret} and \say{secrets} as two different words. For this work, these words are the same. Therefore, word embedding provides a unique approach to encoding the information better. For this problem, we also want the model to generalize well. Our training set is a small fraction of the commands we want to predict. The naming convention of commands spans many teams. So it is vital to capture the information without needing to train on every variation of the word \say{secret}

\section{Tokenizing the CLI}
\topic{Powershell response}
PowerShell commands are combined with response names and types to map this problem to what is known as a document in the NLP space. All three values are in Pascal Case, so we can separate words by the first capital letter, removing non-alphanumeric values. For example, \say{New-AzKeyVault} becomes \say{New Az Key Vault.} Powershell contains several variables with multiple upper-case characters in a row. For this, we split the upper case into one separate word. For example, complicated names like \say{New-AzVMConfig} become \say{New Az VM Config.} The result is that each feature becomes a document with words inside that document.

\topic{Powershell tranformation}
The system for PowerShell is an azure command that has a response. This response is a hierarchal set of variables where each variable has a type and a value \ref{powershell}. For this work, we want to redact or allow each variable individually. Therefore, the information we have for each variable includes 1) variable name, 2) variable type, 3) variable value, 4) variable parent name hierarchy, and parent variable type hierarchy. The variable name is the name of the variable name as assigned by the system. Variable type is the object type, such as int, float, string, or custom object. Finally, the variable value is the actual value returned and set in that variable.

\topic{From Command to Feature}
Finally, to turn it into words and sentences, we implement tokenization on non-alpha numerics and go from lowercase to uppercase. Table \ref{powershell} shows the original PowerShell command. This single command is mapped to 4 items for machine learning with features and labels, as shown in Table \ref{features}. Finally, Table \ref{documents} demonstrates concatenating and tokenization to get something that looks like documents used for NLP.

\begin{table*}
\begin{verbatim}
Command:
Get-AzLocation

Response:
Location: eastasia
DisplayName: East Asia
Providers: {Microsoft.Devices,
    Microsoft.Cache, ...}
\end{verbatim}
\caption{Example PowerShell command}
\label{powershell}
\end{table*}

\begin{table*}
\begin{center}
\begin{tabular}{ c c c c c c }
\hline
Command & Module & Field Name & Field Type & Parent Name & Parent Type \\
\hline
Get-AzLocation & Resources & & PSResource- & & \\
& & & ProviderLocation & & \\
Get-AzLocation & Resources & Location & string & & PSResourceProviderLocation \\
Get-AzLocation & Resources & DisplayName & string & & PSResourceProviderLocation \\
Get-AzLocation & Resources & Providers & System.Collections.- & & PSResourceProviderLocation \\
& & & Generic.List$<$string$>$ & & \\
\hline
\end{tabular}
\end{center}
\caption{Example list of machine learning features}
\label{features}
\end{table*}

\begin{table*}
\begin{center}
\begin{tabular}{ c }
\hline
Document \\
\hline
Get Az Location;Resources;;PS Resource Provider Location;; \\
Get Az Location;Resources;Location;string;;PS Resource Provider Location \\
Get Az Location;Resources;Display Name;string;;PS Resource Provider Location \\
Get Az Location;Resources;Providers;System Collections Generic List string;;PS Resource Provider Location \\
\hline
\end{tabular}
\end{center}
\caption{Conversion of commands to documents}
\label{documents}
\end{table*}

\topic{variable value}
For this work, we omit variable values. There are three reasons we overlook it in model training and prediction. The first is that the variable value can be any type. Therefore it may not be a string or number. Casting all these values to strings may be cumbersome, or setting up a model to handle the native type is difficult. Second, we are predicting whether the variable value is sensitive and do not want this value to egress the Sovereign Cloud. If we include this value in our training set, we limit where and how we can train the model. The training data will have sensitive data, so human interaction with the training data is not allowed. It makes the data more challenging to handle. Finally, the most sensitive data are passwords, which are strings of random characters. It is unlikely that the model will pick that up as a signal to indicate a password as much as a field name of \say{Password} or \say{Passwd.}

\topic{transformation}
We desire to preprocess the data set to produce a similar data set to NLP problems. All the features are text and, if combined, resemble a sentence. In NLP, words are in a spoken or written language. For this work, we use words to mean the combination of letters determined by our tokenization process. Although similar to a sentence, it does not contain correct English grammar. The coding standard for variable names and types in PowerShell is Pascal Case. Pascal Case is where a variable name or type is a combination of words that join into one word, and the first letter is in uppercase while the rest are in lowercase. For example, a Pascal case for the phrase virtual machine name is VirtualMachineName. The initial preprocessing tokenizes between each lowercase letter and an uppercase letter.

\section{NLP Techniques}
\subsection{Bag of Words}
For this work, we explore five techniques for converting text to numbers for use in machine learning models. This technique creates a dictionary based on all the words in the training set. Each vocabulary item has a positional value. The transformation turns each item from a sentence to an array where the array is the vocabulary length, and the word count found in the document is assigned to the positional value.

Equation \ref{tf_equation} shows the term frequency ($tf$) of each term ($t$) per document ($d$).

\begin{equation}
\label{tf_equation}
tf(t, d) =
    \frac{
        f_{t,d},
	}
	{
	    \sum_{t' \in d}{f_{t'd}}
	}
\end{equation}

\topic{bow advantage/disadvantage}
The advantage is that this transformation is easy to implement. It also reinforces items that have words used frequently. One downside of BOW is that document lengths can significantly affect counts. A lengthy document can naturally have more of a single word, even though that may not reinforce its importance in predicting that document. However, this does not significantly affect this problem because all documents are close to the same length. BOW also does not consider the ineffectiveness of frequently used words or a word's relative position.

\subsection{TF-IDF}
TF-IDF resolves some of the BOW disadvantages. This transformation works similarly to BOW but includes a factor that decreases the value of a word-based if used more frequently across all documents. Equation \ref{idf_equation} shows the IDF based on the number of document occurrences of the term ($n_t$) and the number of documents ($N$).

\begin{equation}
\label{n_equation}
N = |D|
\end{equation}

\begin{equation}
\label{nt_equation}
n_t = |\{d \in D: t \in d\}|  
\end{equation}

\begin{equation}
\label{idf_equation}
idf(t, D) = \log{\frac{N}{n_t}}
\end{equation}

\begin{equation}
\label{tfidf_equation}
tfidf(t, d, D) = tf(t, d) * idf(t, D)
\end{equation}

Less frequently used TF-IDF still has the disadvantage of not considering word order.

\subsection{Word Embedding}
Word embedding (WE) is a technique that converts each word to a unit vector with a specified number of dimensions. The mapping of word embeddings results in vectors that are close in Euclidean distance having similar meanings. Since each word in order is converted to a vector, concatenating all the vectors results in a feature set sensitive to word order.

WE creates unique vectors for each word. The size of the vector can be tuned to get the best results. WE algorithm is shown in figure \ref{algorithm:we}. Mikolov, Chen, Corrado, and Dean \cite{mikolov2013efficient} developed the Word2Vec algorithm.

Using a pre-trained Word2Vec model converts words into similar vectors that are not similar. For example, the initial for Azure, \say{Az,} and the initial for the identifier \say{Id} can be misinterpreted as Arizona and Idaho, respectively. A pre-trained Word2Vec model sees these are similar items and therefore assigns them similar vectors. To mitigate this, we trained a Word2Vec using all of the Azure Powershell commands and responses. Since this is more akin to unsupervised learning, we do not need labels. This trained model performs better and learns relationships with this specialized vocabulary in Azure Powershell.

\begin{algorithm}
\begin{algorithmic}[1]
\caption{WE algorithm}
\label{algorithm:we}
    \INPUT max words $m$, vector length per word $l$
    \OUTPUT float matrix $V$
    \STATE \emph{Initialize:} $v_k = 0 \ \forall k \epsilon {1,...,m*l}$, $i = 0$
    \STATE W = vector of words from tokenization of document
    \STATE $J \leftarrow \|W\|$
    \FOR{$j = 1, 2, ... , J$}
        \STATE $w \leftarrow Word2Vec(W_j)$
        \FOR{$k = 1, 2, ... , l$}
            \STATE $v_{i+k} = w_k$
        \ENDFOR
        \STATE $i \leftarrow i + 1$
        \IF{$i > l * m$}
            \STATE END LOOP
        \ENDIF
    \ENDFOR
\end{algorithmic}
\end{algorithm}

Table \ref{we_example} shows a length-3 vector used for word embedding and the euclidean distance of each word from the WE of \say{password.} Using WE, the word \say{passwords} is closer to \say{password} than to \say{private.} This demonstrates an advantage of WE in that similar words have similar vector representations.

\begin{table}[htbp]
\begin{center}
\begin{tabular}{ c c c }
\hline
Word & WE 3 & Distance \\
\hline
password & [0.1, 0.2, 0.5] & 0.00 \\
passwords & [0.2, 0.1, 0.4] & 0.17 \\
private & [0.9, -0.1, 0.1] & 0.94 \\
\hline
\end{tabular}
\end{center}
\caption{WE example}
\label{we_example}
\end{table}

\topic{we disadvantage}
The disadvantage of word embeddings is not being able to handle homonyms. Words spelled the same but with different meanings have the same encoding. Acronyms in variable names and commands cause similar issues. For example, VM could mean \say{virtual machine} for one PowerShell command and \say{virtual memory} for another. In NLP problems, the word order in a sentence can help a model determine the word's meaning. Also, using modeling for parts of speech help in this problem. These are not as helpful in this domain because commands and variable names while being self-descriptive, are not complete sentences, and there is no requirement to use correct grammar. Also, a system like PowerShell, and future CLIs that will be supported, are built by multiple teams without overarching coding standards.

\subsection{BOW Per Feature}
Another approach explored is using BOW per feature (BOW-PF). For example, the word \say{Connection} in the field name or parent field name is more important than the word \say{Connection} found in the command's name. This hybrid approach adds a type of word order sensitivity. For this transformation, we train the BOW transformation on the original data set to use the entire training vocabulary. From there, we use BOW to transform each feature separately. Instead of using the semicolon concatenated document, we use the features from Table \ref{features}. This data creates six times as many features, but now BOW counts are per feature.

\topic{Describe algorithm}
The BOW transform trains on all features as tokenized sentences. Apply the transform to each of the features separately. This results in transformed features of the size: original feature size * vocabulary size (v). The first feature will be the count of the first vocabulary word in the first feature. The second feature will be the count of the second vocabulary word in the first feature. The 1+v feature will be the count of the first vocabulary word in the second feature.

\begin{algorithm}
\begin{algorithmic}[1]
\caption{BOW-PF algorithm}
\label{algorithm:bowpf}
    \STATE \emph{Parameters:} m = max words, C = vector of features each containing a document, n = number of features
    \STATE \emph{Initialize:} $v_{i,j} = 0 \ \forall i \epsilon {1,..,n} j \epsilon {1,...,m*l}$, $i = 0$
    \STATE VOCAB = unique list of words across all features in the data set
    \STATE $J \leftarrow \|C\|$
    \FOR{$j = 1, 2, ... , J$}
        \FOR{$k = 1, 2, ... , n$}
            \STATE $w \leftarrow BOW(c_{j,k}, VOCAB)$
            \FOR{$l = 1, 2, ... , \|w\|$}
                \STATE $v_{j+k} = w_l$
            \ENDFOR
        \ENDFOR
    \ENDFOR
\end{algorithmic}
\end{algorithm}

\error{Make sure we have five sources within the last four years}
\error{Cite more for BT and NN}
\error{Cite BOW research}
\error{Validate algorithms}

\topic{BOW-PF advantage/disadvantage}
This transformation has similar advantages to BOW. Where it differs is the added advantage of separating word count per feature so that the model can learn the nuances of words in command name vs. variable name. However, a disadvantage is that it creates more features and can take an ML model longer to learn. In future work, this can be resolved by adding a feature selection step that could reduce the number of features before training a final model.

\subsection{TF-IDF Per Feature}
Similar to the BOW-PF, the TF-IDF per feature (TF-IDF-PF) trains the transform using the TF-IDF on the documents shown in Table \ref{documents}. Then applies that transform to each feature separately. This process also has six times as many features. The advantages include TD-IDF's ability to scale word importance to word prevalence across documents.

\begin{algorithm}
\begin{algorithmic}[1]
\caption{TF-IDF-PF algorithm}
\label{algorthm:tfidfpf}
    \STATE \emph{Parameters:} m = max words, C = vector of features each containing a document, n = number of features
    \STATE \emph{Initialize:} $v_k = 0 \ \forall k \epsilon {1,...,m*l}$, $i = 0$
    \STATE VOCAB = unique list of words across all features in the data set
    \STATE $J \leftarrow \|C\|$
    \FOR{$j = 1, 2, ... , J$}
        \FOR{$k = 1, 2, ... , n$}
            \STATE $w \leftarrow TF-IDF(c_{j,k}, VOCAB)$
            \FOR{$l = 1, 2, ... , \|w\|$}
                \STATE $v_{j+k} = w_l$
            \ENDFOR
        \ENDFOR
    \ENDFOR
\end{algorithmic}
\end{algorithm}

\topic{Future work for BOW-PF}
In the future, we would like to explore reducing the features used for BOW-PF and TF-IDF-PF. One idea is to use BOW applied to the features as a sentence. Train using a tree model and use feature importance to reduce the vocabulary. Then apply the reduced vocabulary BOW transform to each feature separately. Doing this will decrease training time. It should also help with the generalization of the model. This technique excludes rare features that can make the model overfit.

\section{Experiments}

\subsection{Machine Learning Models}
\topic{Model Intro}
To compare different preprocessing steps, we want to pick a couple of different machine learning models to compare results. Some model types work differently in handling hundreds of features, which causes underfitting or overfitting.

\topic{Logistic Regression, Boosted Trees, Neural Networks}
Logistic Regression (LR) is a prime candidate as a baseline for comparison. It trains quickly relative to other types of models. It is also easier to analyze the relationship between features and the label. A more advanced method is AdaBoost Trees (BT)\cite{freund1997decision}. This method trains multiple trees on the training data, with each subsequent group of trees trained on a weighting based on incorrectly classified examples. The last method used is Neural Networks (NN). Each of these methods can be refined to build optimal F5-Scores. This work aims not to find the best modeling technique but to determine which transformation technique works well in each model type. The same hyperparameters are used across all transformations for each model type to compare transformation methods accurately.

\subsection{Evaluation Data Sets}
We use a manually labeled data set containing over 60,000 entries derived from 1,420 commands. Those commands make up 4\% of all Azure PowerShell commands. In addition, we labeled items that allow for authentication, authorization, or specifically designated sensitive for this data set. This includes connection strings, certificates, passwords, and values stored in a Key Vault. This does not include IP, URLs, or ports. These items aid in the connection to a resource but do not provide authentication or authorization.

\begin{table}[htbp]
\begin{center}
\begin{tabular}{ c c }
\hline
Name & Count \\
\hline
Azure Command Coverage & 1,420 \\
Training size & 62,579 \\
Redacted  & 2,155 \\
\hline
\end{tabular}
\end{center}
\caption{Training data statistics}
\label{training_data}
\end{table}

Some libraries have more redacted fields.

\topic{experiment setup, number of runs}

\subsection{Metrics}
\topic{metric model}
The model should attempt to redact all sensitive fields. It is understood by the customer and us that this is impossible, whether it be because of statistically based models that could miss some sensitive information or human mistakes. To that end, we work with the customer to provide the best solution and work to asymptotically approach 100\% recall in partnership with the customer. False positives are not a major issue. If something is redacted that should not be, it could hinder the DRI's work. False negatives are a bigger issue. They allow DRIs to see sensitive information. Sensitive information is defined as passwords, connection strings, and other objects that are keys to information or keys to resources. For this work, we are currently assuming DRIs as good users, and the system is reducing information to just a need-to-know basis.

Therefore we need a metric that encapsulates the critical importance of recall. We must balance that will precision. If too many values are hidden, it is difficult for a person to do maintenance. Therefore we have to balance security with usefulness. A house without windows or doors is the most secure but also difficult to live within.

The best option is to use the $F_\beta$ score shown in Equation \ref{fbeta_equation}. We want a continuous function that takes into account a tradeoff between precision and recall. This is a practical solution but does not provide a single metric if we want to increase recall. Therefore one of the best single metrics for this work is the $F_\beta$ score. This equation specifies a tradeoff relationship between recall and precision. For this work, we define that recall is five times more important than precision. For an f-score to be better, if we increase recall by one percent, that is worth 5\% false positives (decrease in precision.

\begin{equation}
\label{fbeta_equation}
F_\beta = (1+\beta^2) *
    \frac{
        precision * recall,
	}
	{
	    (\beta^2 * precision) + recall
	}
\end{equation}

Each model has a range of $F\beta$-scores determined by a score threshold. This demonstrates the differences between the models based on these curves.

Because models are statistically derived, often taking samples of training data to improve the loss, we run each model 20 times and take the average max f5-score. With this, we also run statistical significance.

\topic{infrastructure metrics}
For this work, we propose several metrics for comparison. One often-used metric is the area under the ROC curve (AUC). This is the area under the receiver operator curve (ROC). ML models output probabilities of an event. The split of this probability defines one of two categories based on a threshold for a two-categorical system. Over that value, we select one. Under we select the other. As the threshold changes, the metrics change. AUC is a single metric across all the threshold's possible values. It tends to indicate whether a model has better overall thresholds than another. AUC is a valuable metric in some problem domains. However, sometimes, our data is highly skewed. As a result, AUC can yield misleading results.

For our work, we are most interested in precision and recall. Sensitive data should never leave the Sovereign Cloud. For that reason, recall is critical. For comparing models, we want to hold recall at 0.99 or higher and continue to increase precision. This number means that the model redacts sensitive information 99\% of the time at the cost that it could block nonsensitive data, which may be helpful for a DRI to do their work. Therefore, we want to maintain a high standard of 0.99 or higher and continually improve at not blocking nonsensitive data.

\subsection{Vocabulary Size}
Each of the transformations discussed uses a vocabulary of words. In NLP, words are a combination of letters that make a word in a specified spoken or written language. We refer to a combination of letters as a word for this work, although acronyms and shortened words are frequently used in programming. Therefore our vocabulary differs a little from words found in the English language.

The vocabulary or number of unique words in the training set determines the number of features for BOW and TF-IDF. The features for BOW are the counts of each vocabulary word defined by equation \ref{tf_equation}. The features of TF-IDF are the same in the count but determined by equation \ref{tfidf_equation}. 
For WE, fixed-length vectors represent each word. The average number of words in the tokenized command data set is 78. In order to make a fair comparison to the other transformations, the size of the WE final vector is equal to the vocabulary size of BOW and TF-IDF. This experimentation uses a length of 20 float vectors per word.

\subsection{Feature Importance}
Table \ref{bow_features} and \ref{tfidf_features} display the differences in the importance of different vocabulary words using BOW and TF-IDF transform. One crucial difference and something important to this problem is that \say{certificate} and \say{certificates} are treated as two separate words. For BOW, the word, certificate, is a rarer word, so it is not as crucial as \say{certificates.} On the other hand, TF-IDF emphasizes the importance of the IDF multiplier and makes \say{certificate} almost as important. WE algorithm handles this by making the vector encoding of the two words very close in Euclidean distance. As a result, each algorithm does a little better at emphasizing the importance or lack of difference between \say{certificate} and \say{certificates.}

\begin{table}[htbp]
    \begin{center}
    \begin{tabular}{ c c }
    \hline
    Transform & Feature Size \\
    \hline
    BOW & 1,559 \\
    TF-IDF & 1,559 \\
    WE & 1,560 \\
    BOW-PF & 9,354 \\
    TF-IDF-PF & 9,354 \\
    \hline
    \end{tabular}
    \end{center}
    \caption{Feature size}
    \label{feature size}
\end{table}

\begin{table}[htbp]
    \begin{center}
    \begin{tabular}{ c c }
    \hline
    Word & Importance \\
    \hline
    key & 0.071 \\
    certificates & 0.041 \\
    encryption & 0.032 \\
    secret & 0.032 \\
    network & 0.029 \\
    name & 0.022 \\
    vault & 0.021 \\
    compute & 0.021 \\
    management & 0.020 \\
    string & 0.019 \\
    \hline
    \end{tabular}
    \end{center}
    \caption{Feature importance for BOW}
    \label{bow_features}
\end{table}

\begin{table}[htbp]
    \begin{center}
    \begin{tabular}{ c c }
    \hline
    Word & Importance \\
    \hline
    certificates & 0.049 \\
    key & 0.042 \\
    certificate & 0.041 \\
    name & 0.037 \\
    string & 0.032 \\
    osprofile & 0.031\\
    az & 0.030 \\
    encryption & 0.024 \\
    system & 0.023 \\
    id & 0.022 \\
    \hline
    \end{tabular}
    \end{center}
    \caption{Feature importance for TF-IDF}
    \label{tfidf_features}
\end{table}

BOW-PF introduces a partial word order. It weighs words used in the field name and parent name more strongly than the module name or the actual command. The feature importance of BOW-PF is shown in table \ref{bowpf_features}. Interestingly, the model weights the word key high in the field name.

\begin{table}[htbp]
    \begin{center}
    \begin{tabular}{ c c c }
    \hline
    Feature & Word & Importance \\
    \hline
    Field Name & key & 0.0562 \\
    Parent Type & network & 0.0343 \\
    Field Name & data & 0.0253 \\
    Field Name & certificate & 0.0239 \\
    Parent Name & network & 0.0208 \\
    Field Name & name & 0.0206 \\
    Parent Type & service & 0.0188 \\
    Field Name & string & 0.0186 \\
    Field Type & encryption & 0.0184 \\
    Parent Type & psmanaged & 0.0184 \\
    \hline
    \end{tabular}
    \end{center}
    \caption{Feature importance for BOW-PF}
    \label{bowpf_features}
\end{table}

TF-IDF-PF has a similar finding as TF-IDF, where the plural of a word has similarly high importance, shown in table \ref{tfidfpf_features}.

It is also worth noting that \say{string} is an essential word in all models. Almost all secrets come from strings or complex string types. Very little sensitive information is stored in integers, booleans, or enumerated types.

\begin{table}[htbp]
    \begin{center}
    \begin{tabular}{ c c c }
    \hline
    Feature & Word & Importance \\
    \hline
    Field Name & key & 0.0630 \\
    Field Type & collections & 0.0308 \\
    Command & az & 0.0283 \\
    Field Name & data & 0.0280 \\
    Parent Type & osprofile & 0.0252 \\
    Field Type & azure & 0.0212 \\
    Parent Type & models & 0.0211 \\
    Field Type & system & 0.0203 \\
    Parent Type & cluster & 0.0190 \\
    Field Name & keys & 0.0189 \\
    \hline
    \end{tabular}
    \end{center}
    \caption{Feature importance for TF-IDF-PF}
    \label{tfidfpf_features}
\end{table}

\topic{training}
The training data is built by manually labeling 1500 commands and each of the response attributes. The data is split into 80\% training and 20\% validation data. The split is done by Azure Powershell command so that the validation set has commands that have not been used in training. This should yield a good estimate of the generalizability of the algorithm and how well it works on unseen commands.

\begin{figure*}%
\centering
\subfigure[LR]{
    \includegraphics[width=5cm]{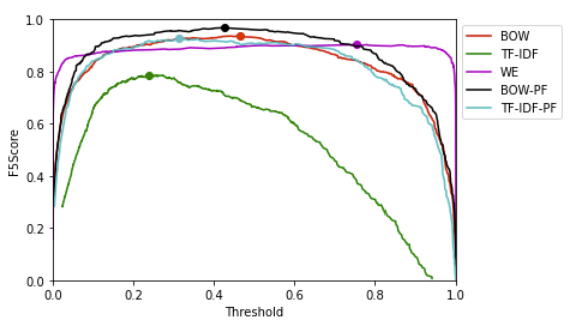}
    \label{lr}
}
\subfigure[BT]{
    \includegraphics[width=5cm]{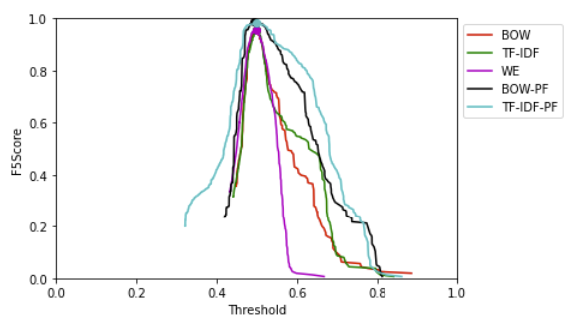}
    \label{ada}
}
\subfigure[NN]{
    \includegraphics[width=5cm]{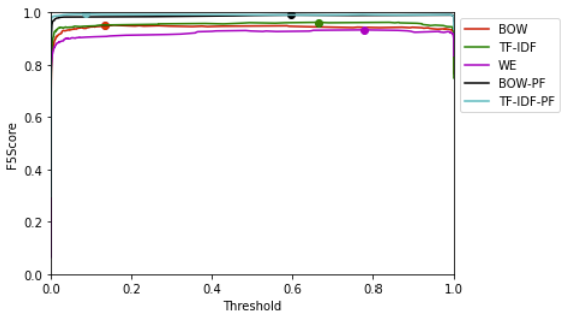}
    \label{nn}
}
\subfigure[LR zoomed]{
    \includegraphics[width=5cm]{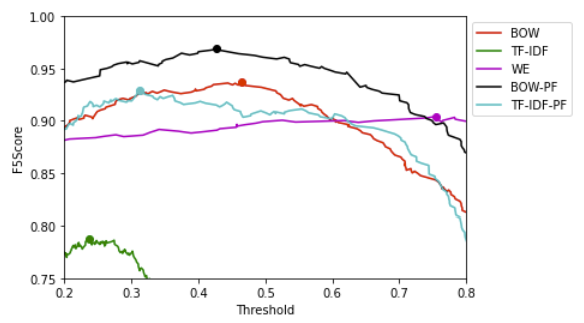}
    \label{lr_zoom}
}
\subfigure[BT zoomed]{
    \includegraphics[width=5cm]{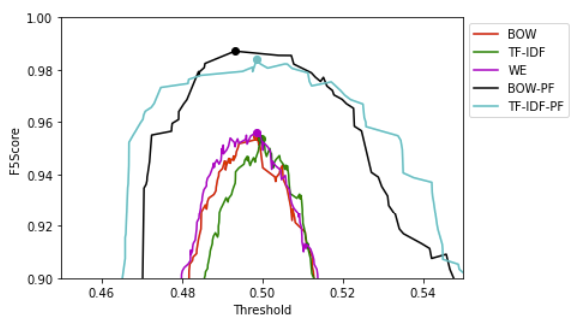}
    \label{ada_zoom}
}
\subfigure[NN zoomed]{
    \includegraphics[width=5cm]{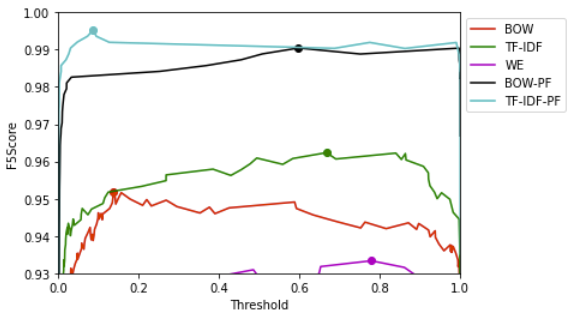}
    \label{nn_zoom}
}
\caption{F5-Score for transforms and models.}
\label{f5score}
\end{figure*}

\topic{how we run}
Each model is run 20 times, with the results averaged. Displayed in figures \ref{lr} are the runs of each transform. Some depict using logistic regression, and others depict using boosted trees. First, each model trains using the training set. Next, calculating the maximum F5-score is done using the test set. Finally, using the trained model and the threshold for the max F5-score, the validation set is used to determine F5-score, as seen in new examples.

\subsection{Metric Comparison}

\topic{Describe max F5-score}
$F_\beta$-score changes with the threshold selected. This model must make a binary decision, so we choose the threshold to give us the max F5-Score. We use the test data set to determine the threshold with the max F5-Score. We then use the validation set to test the model and threshold against unseen examples to determine how well the model will generalize and how well it will work in a production setting.

\topic{Describe what is in each curve}
Each curve shows the F5-Score, as calculated from equation \ref{fbeta_equation}, for all thresholds. The point on the curve is the max F5-Score overall threshold value. It should be noted that a curve with a plateau top has a robust threshold. Slight changes in threshold do not change F5-Score much. It also implies that unseen examples are not sensitive to slight changes in threshold value. A sharp peak at the maximum F5-Score implies an unstable maximum. Using this model against this or other data sets could result in very different F5-Scores if anything is slightly perturbed.

\topic{Describe LR Results}
For the logistic regression example, TF-IDF is the weakest. BOW is effective, while BOW-PF is the most effective. The PF is a form of word importance that weights the words. TF-IDF does not work as well in highly skewed data. The word frequencies help predict both the positive and negatively labeled items. Infrequent words help identify risky items, but most items are not risky, so words that help predict non-risky items are frequent, and therefore TF-IDF decreases the weights.

\topic{Describe BT Results}
BT has more of a peak to most curves. Unlink LR BT is discontinuous. The inference of a BT model with perturbation of features causes different parts of the tree to be traversed. The ordering of worst-to-best algorithms stays relatively consistent across models. BT's ability to handle discontinuous functions allows it to fit the data better, resulting in a higher F5-Score. BOW-PF still performs the best as it uses word counts, does not discount frequently used words, and includes a type of word ordering captured in the Per-Feature part of the algorithm.

\topic{Describe best algorithm and advantages}
This work demonstrates how BOW-PF and TF-IDF-PF can be used to apply a word order that is not in BOW or TF-IDF. It is recommended that if you use LR or BT, BOW-PF is the transformation that should be used. BOW-PF captures the use of words in each feature, for example, emphasizing the field name having the word \say{key} over the command having the word \say{key.} The inverse of the document frequency is not handled well in LR and BT. NN, on the other hand, picks up the nuances of the inverse of document frequency and handles TF-IDF and TF-IDF-PF better than the BOW counterparts.  

\begin{table}[htbp]
    \begin{center}
    \begin{tabular}{ c c c c }
    \hline
    Transform & LR & BT & NN \\
    \hline
    BOW & 0.937 & 0.955 & 0.952 \\
    TF-IDF & 0.788 & 0.954 & 0.962 \\
    WE & 0.904 & 0.956 & 0.933 \\
    BOW-PF & 0.969 & 0.987 & 0.990 \\
    TF-IDF-PF & 0.929 & 0.984 & 0.995 \\
    \hline
    \end{tabular}
    \end{center}
    \caption{Max F5-Scores for transforms}
    \label{f5_scores}
\end{table}

\section{Conclusion and Future Work}
\topic{What was accomplished}
This work demonstrates how to convert PowerShell commands to the well-researched NLP domain. From there, we compare popular techniques and find that BOW-PF performs best at creating features for different types of models. This work exploited the sentence-like structure of commands, variable names, and variable types. Using F5-Score as a metric for comparison allows us to summarize system goals into one value for optimizing across thresholds and in subsequent retraining.

\topic{Expanding to other commands}
Slight adaptions to this work yield a system that works across different CLIs. Systems other than PowerShell will be more complex because the output is not in a structured programming object. Also, the variables' names are Pascal-case for PowerShell, but other CLIs will require different transformations to convert commands and variables to words. This will require more work to convert as even if the responses are structured as objects and the variables' names are Pascal-case, the names are variables are still created by different development teams. It is similar to the NLP space of handling different languages. For example, one CLI might call a password variable \say{Password,} whereas another CLI interface might call it \say{Psswd.} The system will need to be trained to pick up new vocabulary.

\topic{Expanding to APIs}
This work can extend to APIs. Powershell, behind the scenes, makes API calls to pass commands and receive back the response. Powershell turns those JSON or XML responses into objects. This system is built on structured response objects. APIs have similar structures in which there is an endpoint, similar to a command, and a response, often a JSON object, analogous to the response object of PowerShell. Therefore with little work, the risk modeling can be moved to API space to redact or mask values returned in API calls.

\newpage 
\bibliographystyle{ieeetr}
\bibliography{bib}

\end{document}